\documentclass[prl,aps,superscriptaddress,12pt]{revtex4}
\usepackage{graphics}
\begin{document}          

\title{Tailoring the profile and
interactions of optical localized structures }

\author{P.L. Ramazza}
\affiliation{Istituto Nazionale di Ottica Applicata, I50125 Florence-Italy}
\author{E. Benkler}
\affiliation{Institute of Applied Physics, Darmstadt}
\author{U. Bortolozzo}
\affiliation{Istituto Nazionale di Ottica Applicata, I50125 Florence-Italy}
\author{S. Boccaletti}
\affiliation{Istituto Nazionale di Ottica Applicata, I50125 Florence-Italy}
\affiliation{also Universidad de Navarra, Pamplona, Spain}
\author{S. Ducci}
\affiliation{Ecole Normale Sup\'erieure de Cachan, 94235 Cachan Cedex, France}
\author{F.T. Arecchi}
\affiliation{Istituto Nazionale di Ottica Applicata, I50125 Florence-Italy}
\affiliation{also Universita' di Firenze,  Italy}

\begin{abstract}
We experimentally demonstrate the broad tunability of the main
features of optical localized structures (LS) in a nonlinear
interferometer. By discussing how a single LS depends on the
system spatial frequency bandwidth, we show that a modification of its tail
leads to the possibility of tuning the interactions between LS
pairs, and thus the equilibrium distances at which LS bound states
form. This is in  agreement with a general theoretical model
describing weak interactions of LS in nonlinear dissipative
systems.

\end{abstract}

\maketitle

Localization of spatial patterns is a subject of major current
interest in the research on nonlinear dissipative dynamical
systems. The studies about this topics have naturally followed and
sided those dedicated to the formation of temporal and spatial
solitons in Hamiltonian systems \cite{solitoniHamilton}. 
Analytical and numerical works have identified several distinct
mechanisms leading to structure localization in dissipative
systems \cite{Riecke}, and  experimental observations of this phenomenon
have been recently offered in several systems, such as fluid
dynamics \cite{solitonifluidi}, chemistry \cite{solitonichimici},
granular materials \cite{solitonigranulari} and  nonlinear optics
\cite{solitoniottici}.

In particular, optical localized structures (LS), to which we will
also refer to as {\it dissipative solitons} in the following, are
objects of intense research, also in view of possible applications
as pixels in devices for information storage or processing. So
far, the existence of optical dissipative solitons has been
theoretically predicted in many passive \cite{Tlidi} and
active \cite{sistemiattivi} configurations, and optical LS have
been observed in photorefractive cavities \cite{AndersonWeiss1}
and in passive nonlinear interferometers, based either on the
"thin slice with feedback" scheme
\cite{DarmstadtLS,MuensterLS,NoiLS}, or on a microresonator filled
with a semiconductor medium \cite{KuscelewitzTredicce}. More
recently, the interactions between LS have been shown to give rise
to the formation of a discrete set of bound states
\cite{MuensterLS}.

To our knowledge, very little is known about the dependence of the
LS's features on the experimental parameters. The present work
addresses this issue, by investigating how the spatial frequency
bandwidth of a nonlinear interferometer can be utilized to tune
both the spatial profile of each single soliton, and the
interaction forces occurring between two of them. A quantitative
experimental evidence is given of the crucial role played by the
oscillatory tails of a single LS in determining the interaction forces
between solitons.

Our experimental system consists of a Liquid Crystal Light Valve
(LCLV) closed in an optical feedback containing both
interferential and diffractive processes. When
an initially plane wave is sent into the system, its phase
$\varphi (\vec r, t)$ evolves according to \cite{NoiLS}

\begin{equation}
\tau {\partial \varphi \over \partial t}=-(\varphi -\varphi _0) + l_d^2 \nabla^2_{\perp}
\varphi+ \alpha  I_0 \mid e^{{-il\nabla^2}\over {2k_0}} (Be^{-i\varphi} +C) \mid ^2
\label{uno}
\end{equation}

where $\varphi_0=\pi$ is the phase working point of the LCLV, and
$\tau$ and $l_d$ are its response time and diffusion length
respectively.
The source term in the right hand side of
Eq. (\ref{uno}) depends on the free propagation length $l$ in the
feedback loop, as well as on the laser light wavenumber $k_0$ and
on the parameters $B$ and $C$, that tune the relative weight of
diffraction and interference in the system. Finally, $I_0$ is the
incident laser intensity, and $\alpha$ describes the Kerr-like
response of the LCLV. Here, $B = \cos^2 \theta$, $C = \sin^2
\theta$, where $\theta$ is the (experimentally adjustable) angle
between the director of the nematic liquid crystals of the LCLV
and the transmissive axis of a polarizer oriented along the
polarization direction of the incident light.

In a previous work \cite{NoiLS}, we have characterized the state
diagram of the interferometer in the parameter plane
$(\theta,I_0)$, finding that localization of patterns occurs for a
broad range of $\theta$ values ($\simeq 35^o$ to $58^o$). This
phenomenon is related to the presence of a subcritical
bifurcation, connecting a lower uniform branch to an upper
patterned one. In these  conditions, the formation of isolated
spots connecting the two branches is typical \cite{Tlidi,sistemipassivi,Aranson}.
Besides $\theta$ and $I_0$, the scenario of observable patterns
crucially depends on the spatial frequency bandwidth $q_B$ of the
interferometer, which can be experimentally controlled by means of
a variable aperture put in a Fourier plane. In what follows we
discuss the main LS features that emerge by keeping
fixed $\theta =42^o$, and varying $I_0$ and the adimensional
parameter $q_b \equiv {q_B}/{q_{\text{diff}}}$ obtained by
normalizing the system bandwidth to the diffractive
interferometer wavenumber $q_{\text{diff}} = \sqrt{\frac{\pi
k_0}{l}}$.

A first point of interest is to establish the range of existence
of LS in the ($q_b,I_0$) plane. In Fig. 1 we plot the state
diagram of the system in this parameter plane, together with some
snapshots representative of the observed patterns.  All
the experiments are performed at incident laser wavelength
$\lambda = 632$ nm and for $l = 250$ mm. This results in a scale
of the observed patterns of the order of $2 \pi / q_{\text{diff}}
\simeq 0.5$ mm.

Looking at Fig.1, one easily realizes that the range of existence
of LS is very broad, not at all limited to some particular
parameter choices. The lower threshold for the existence of LS
increases for decreasing $q_b$. This is a consequence of the fact that
LS have an internal structure conatining both low and high
frequency components, as it will appear evident in the following.
Therefore, any bandwidth limitation perturbs the LS structure, and
increases the threshold for their existence. At very low $
q_b$ and high intensities, localization of structures is lost and
regular  hexagons are observed, due to the long range correlation
imposed to the pattern by the small bandwidth.

If $I_0$ is kept fixed at high values while $q_b$ is increased,
hexagonal patterns evolve into a space-time chaotic (STC) regime.
The boundary line between STC and LS occurs at decreasing 
intensities when $q_b$ is increased. This indicates that the
regime here generically referred to as STC can arise either from a
strong excitation of a relatively small band of wavenumbers, or
from a weak excitation of a large set of interacting spatial
modes. The indetermination of the boundaries between the different
regimes is of the order of 10 \%. It must be also specified that
the placement of the boundaries depends on the evolutionary
history of the parameters, since we are in presence of a
subcritical bifurcation. The continuous lines in Fig. 1 were obtained by 
decreasing the input intensity, the dashed line by increasing it. 
Localized structures are not observed in this last case.

Scanning the parameters within the domain of LS's existence leads
to sensible modifications in the shape of each structure. In Fig.
2 we show the variation in the LS intensity profile observed by
keeping $I_0$ close to the lower threshold for LS existence and
increasing $q_b$. It is seen here that each structure is formed by
a central peak, 
and by a set of concentric rings forming a tail that shows spatial
oscillations of decreasing amplitudes for increasing distances
from the LS center. The width of the central peak can be roughly 
evaluated as the diameter of the first dark ring in each frame, and appears 
to be practically independent on $q_b$.

The length scale of
the oscillations on the tails is instead strongly dependent on $q_b$. 
Namely, this scale decreases for increasing $q_b$ until $q_b \simeq 3$, and
then saturates to a constant value.

The set of our observations indicates that LS have a  "natural"
unperturbed shape like that displayed for $q_b \geq 3$. By
constraining the system to a bandwidth smaller than this value,
one is then able to tune the LS profile, imposing oscillations  on
the tails at a frequency different from the natural one. The
occurrence of oscillatory tails on LS have been reported in other
physical  systems \cite{Aranson,tails1}, and it is
considered to be a typical signature of  the formation of LS via
pinning of the fronts connecting the uniform and the patterned
states \cite{Riecke}.

The observed LS closely resemble those reported in Ref.
\cite{Aranson}, in which a subcritical real Swift-Hohenberg (S-H)
equation is studied analytically and numerically. This is not
surprising, since our experiment displays a subcritical
bifurcation of a real order parameter to a  patterned state, and
therefore is appropriately modeled by an order parameter equation
of that kind. 
We do not expect that the S-H model describes faithfully all the details 
observed in the experiiment, however. It is known \cite{Neubecker}, for example, 
that the "thin slice with feedback" model, of which our experiment is an 
implementation, presents instabilities at multiple wavenumbers given by 
$q_{\text{N}} = \sqrt{N} q_{\text{diff}}$, $N=1,5,9,...$. Though the highest wavenumbers 
become active at high values of pump parameter due to diffusion, it may be 
expected that they play some role in determining the fine features of the 
LS's.
Our aim in comparing the experimental findings wit the prediction of the 
S-H model is indeed to investigate wheter some fundamental features of the 
observed phenomena can be described in terms of this very general model.

Using the Swift-Hohenberg model, it is found analytically that the LS tails
are described by single spatial scale oscillations, embedded in an
exponential envelope that departs from the lower uniform state.

Though the LS tails in our case display some deviations from the
above ideal behavior, the qualitative agreement between our
observations and the results of the general theory reported in
Ref. \cite{Aranson} is satisfactory. In particular, it is possible
to identify for each value of $q_b$ a dominating spatial scale in
the oscillatory tails. To this purpose, we measure the distance
between successive maxima of a single LS and average this quantity
over all observed maxima. This way, we obtain the dominant spatial
frequency of the tail oscillations, which is then normalized to
$q_{\text{diff}}$ and reported as $q_{\text{tails}}$ in Fig. 3.
The error bars correspond to the measured frequency fluctuations
from the $q_{\text{tails}}$, reflecting the fact that the tail
oscillations are not rigorously at a single spatial scale. Looking
at Fig. 3, one easily realizes that $q_{\text{tails}}$ practically
coincides with $q_b$ for $q_b \leq 3$. At higher values of $q_b$,
no variations in $q_{\text{tails}}$ as well as in  overall LS's profile are
observed.

The shape of the tails is responsible for the interactions between
localized structures. Namely, while for monotonically decreasing
tails, one would expect only attractive or repulsive forces
between LS,  oscillatory tails induce oscillatory signs of the
interactions, thus producing both attractive and repulsive
forces, depending on the distance between the centers of a pair of
LS's \cite{Aranson,tails1}. A recent work \cite{MuensterLS}
has experimentally demonstrated the existence of  a discrete set
of LS bound states, occurring in the presence of oscillations on
the LS tails. 
The selection rule for the discrete set of bound states observed 
has been there put in relation with the spacing of their rings originated 
by diffraction around the central peak.
In the following we show how these bound states can
be in fact tuned by varying the spatial frequency  bandwidth of the
interferometer, and we discuss how the selection of the observed bound 
ststes can be put in the very general framework of a subcritical S-H model.

In Fig. 4 we display a set of different bound states observed for
$q_b$ = 3.6. We notice that the states form a set that can be
ordered following a precise rule, given by simply counting of the
number of maxima and minima that occur along the segment
connecting the two LS centers. We will call this number $n$ as
{\it bound state order} number. Such a feature is encountered for
all values of $q_b$. At small system bandwidths, however, we
observe only the first two or three bound states, instead of the
entire set shown in Fig. 4. This is probably due to the fact
that the binding energy of each state varies with $q_b$, and in
some cases it is not sufficient to keep the LS pair tightly bound
in the presence of unavoidable system inhomogeneities and
fluctuations.

A theory for the interaction of LS pairs was given in Ref.
\cite{Aranson} in the context of  study on a real Swift-Hohenberg
equation. As already discussed, we expect this model to be closely
applicable to our system in the present conditions. Following that
approach, the weak interactions between  a pair of LS with tails
decaying with a length $\mu$ and oscillating at a frequency $\nu$,
lead to a time evolution for the distance $R$ between the LS
centers ruled by

\begin{equation}
{dR \over dt} =  {1 \over R} {d \over dR} (e^{-\mu R} cos(\nu R))
\equiv F(R,\mu,\nu).
\label{due}
\end{equation}

As a consequence, an infinite number of stable bound states are
possible, corresponding to the solutions $F=0$, $dF/dR <0$
of Eq. (2). 
In the limit in which the scales $\mu$ and $\nu$ are
well separated, the difference $R_{n+1}-R_n$ between the
separation distances of two successive bound states corresponds
approximately to the tail oscillation length ${\nu}^{-1}$ of a
single LS. We recall that by {\it weak interaction} we mean a
regime in which the intensity amplitude of one LS is small in the
space region in which the intensity amplitude of the other is
large. In the case of our experimental data, this is true for all
the bound states observed, with the possible exception of the 
lowest order one.

If we assume that Eq. (2) describes correctly the bound state
selection rule in  our experiment, it immediately follows that tuning
of the equilibrium distances should be possible by varying the
scale of the oscillations on the tails of each single LS. In order
to check this point, we measured the quantities $\Delta_{n,n+1} =
R_{n+1}-R_n $, and then averaged them over the bound state order
number $n$. The resulting quantity $\bar d$ (normalized to the
length $\Lambda_{\text{tails}}$) is reported vs. $q_b$ in Fig. 5.
A constant value of the ratio ${{\bar d} \over
\Lambda_{\text{tails}}} \simeq 1$ is observed within the errors,
indicating that the above discussed relation between the
oscillations on the tails of each LS and the selection rule of
bound states is verified. This marks the fact that tuning of the
equilibrium distances between LS's in bound states can be
quantitatively performed in our experiment.

In conclusion, we have given a quantitative evidence of the tuning
of the LS spatial profile in a nonlinear optical interferometer,
using the system spatial frequency bandwidth as a control
parameter. We have discussed the role of the  oscillations
occurring on each single LS tail in determining the interactions
between different LS's. Finally, we have verified the agreement
between the selection rules for the formation of bound states
observed in our experiment, and those  predicted for the same
phenomenon by a general model for pattern formation  in
nonequilibrium systems.

\newpage
\ \
\vspace{4cm}

\begin{figure}[!h]
\begin{center}
\leavevmode
\rotatebox{0}{\resizebox{95mm}{!}{\includegraphics{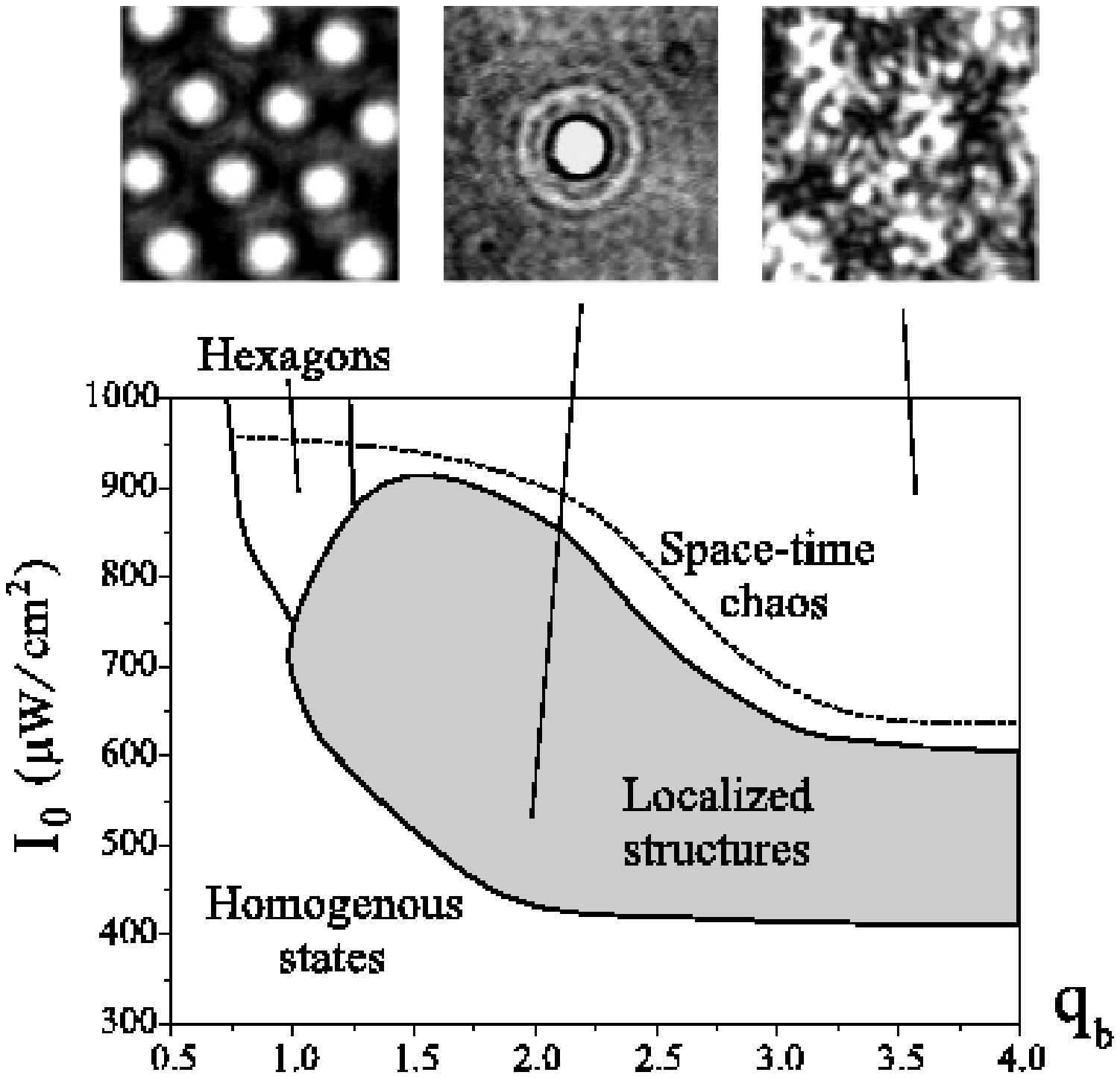}}}  
\end{center}
\caption{State diagram of the system in the ($q_b, I_0$) parameter
plane. Notice that LS emerge for a broad range of parameters (gray
area in the plane). The three reported patterns are snapshots of
the observed hexagons, LS, and space time chaotic states.}
\label{fig1}
\end{figure}

\newpage
\ \
\vspace{2cm}

\begin{figure}[!h]
\begin{center}
\leavevmode 
\rotatebox{0}{\resizebox{70mm}{!}{\includegraphics{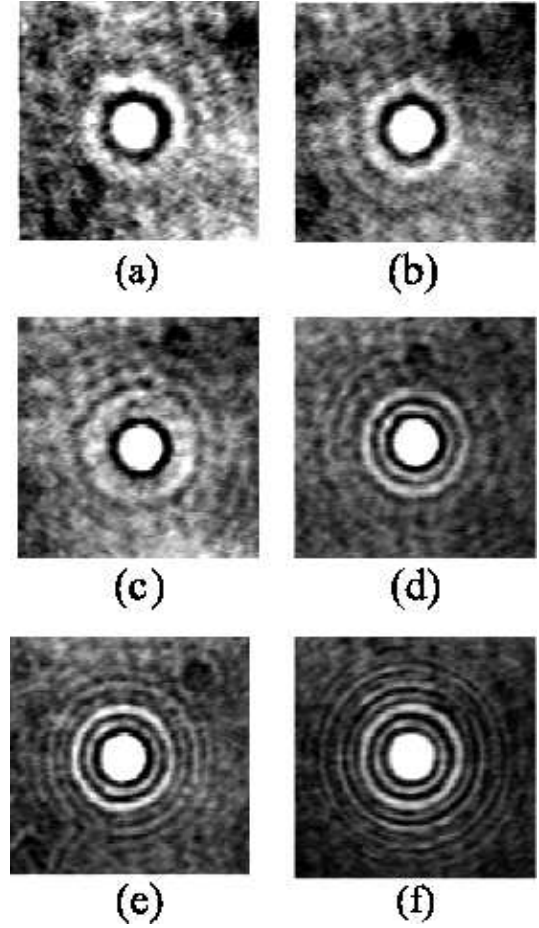}}}  
\end{center}
\caption{Variation of the LS's shape with the system bandwidth.
Snapshots of the observed solitons for (a) $q_b=1.0, I_0=700 \ \mu W/cm^2$ 
(b) $q_b=1.2, I_0=620 \ \mu W/cm^2$, (c) $q_b=1.6, I_0=520 \ \mu W/cm^2$, 
(d) $q_b=2.2, I_0=480 \ \mu W/cm^2$,
(e) $q_b=2.8, I_0=460 \ \mu W/cm^2$, (f) $q_b=4.0, I_0=460 \ \mu W/cm^2$.}
\label{fig2}
\end{figure}

\newpage
\ \
\vspace{4cm}

 \begin{figure}[!h]
 \begin{center}
 \leavevmode 
\rotatebox{0}{\resizebox{120mm}{!}{\includegraphics{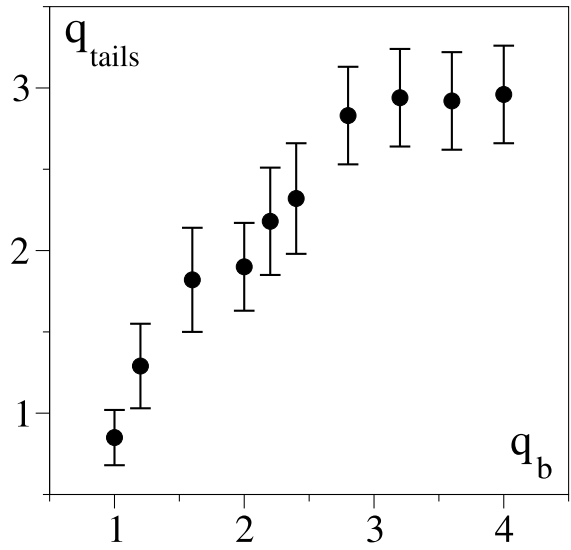}}}  
 \end{center} 
\vspace {-1 cm}
 \caption{Variation in the main frequency of the LS tail oscillations as a function
 of the system bandwidth. Both $q_{\text{tails}}$ and $q_b$ are adimensional quantities
 (see text for definition).}
 \label{fig3}
 \end{figure}

\newpage
\ \
\vspace{1cm}

 \begin{figure}[!h]
 \begin{center}
 \leavevmode 
\rotatebox{0}{\resizebox{40mm}{!}{\includegraphics{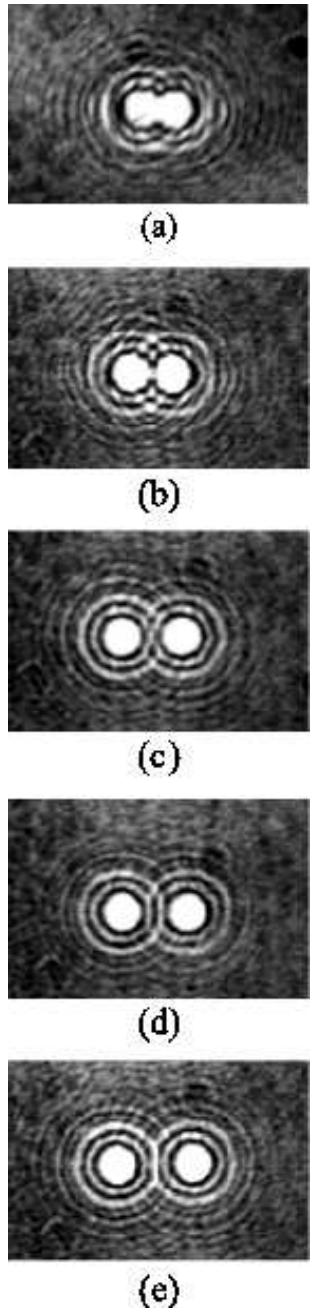}}}  
 \end{center}
 \caption{Snapshots of different bound states observed at $q_b=3.6, 
I_0=500 \ \mu W/cm^2$.
 All patterns (a-e) are obtained by inducing a pair of LS with an
 increasing initial distance between centers, and letting the system evolve
 up to the time at which the stationary bound state is realized.}
 \label{fig4}
 \end{figure}

\newpage
\ \
\vspace{1cm}

\begin{figure}[!h]
\begin{center}
\leavevmode 
\rotatebox{0}{\resizebox{120mm}{!}{\includegraphics{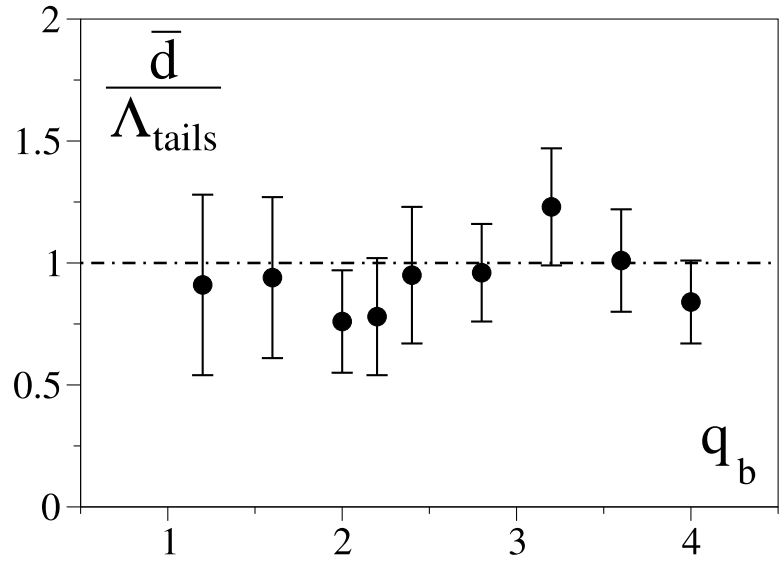}}}  
\end{center}
\caption{${{\bar d} \over \Lambda_{\text{tails}}}$ vs. $q_b$ (see
text for definitions). Both quantities are adimensional. Notice
that, for all measurement a constant value of ${{\bar d} \over
\Lambda_{\text{tails}}} \sim 1$ is realized within the
experimental errors.} \label{fig5}
\end{figure}

\end{document}